  \documentstyle[12pt]{article}
  \documentstyle{fleqn}
\begin{document}
  \newcommand{\barra}[1]{\sopra{#1}{/}}
  \newcommand{\beqn}{\begin{eqnarray}}
  \newcommand{\eeqn}{\end{eqnarray}}
  \newcommand{\beq}{\begin{equation}}
  \newcommand{\eeq}{\end{equation}}
  \newcommand{\partialmu}{\mu {\partial \over {\partial \mu}}}
  \newcommand{\parz}[1]{{\partial \over {\partial #1}}}
  \newcommand{\bra}[1]{ {<\,#1\,|}}
  \newcommand{\ket}[1]{{|\,#1\,>}}
  \newcommand{\ff}{{\bf\tilde{1}}}
  \newcommand{\color}[1]{\left( \hbox{cf} - \hbox{ca}\frac{1}{#1} \right)}
  \begin{titlepage}
  \leftline{\_\hrulefill\kern-.5em\_}
  \vskip -4 truemm
  \leftline{\_\hrulefill\kern-.5em\_}
  \centerline{\small{ 9/92\hfill\#HUTP-92/A040\hfill}}
  \vskip -2.7 truemm
  \leftline{\_\hrulefill\kern-.5em\_}
  \vskip -4 truemm
  \leftline{\_\hrulefill\kern-.5em\_}
  \vskip1.5truecm
  \centerline{\bf QCD non-leading corrections to weak decays in the
  dimensional scheme} \vskip.7truecm
  \centerline{${\rm Giuseppe\:Curci,}^{\,(1)\,}$}
  \vskip1truemm
  \centerline{ ${\rm Giulia \:
  Ricciardi}^{\,(2,\, 3)\,}$}
  \vskip1truecm
  \centerline{\footnotesize{(1) I.N.F.N., sez. di Pisa,
  Via Livornese 582/a, 56010 S. Piero a
  Grado (Pisa) Italy}}
  \vskip1truemm
  \centerline{\footnotesize{(2) Lyman Laboratory
  of Physics, Harvard University, Cambridge, MA 02138, USA}}
  \vskip1truecm

  \begin{abstract}
  We compute the next to leading QCD corrections to weak four fermion
  interactions introducing a scheme that does not require an explicit
  definition of $\gamma_5$ in $d$ dimensions.
  This scheme reduces greatly the difficulties in calculating the  two
  loop anomalous dimensions; we recover the results obtained in the
  previous literature.
  \end{abstract}

  \vskip 1.4in

  \footnotesize{ (3) Work supported by I.N.F.N.-Italy
  and  by the grant of NSF $\#$  Phy 87-14654}

  \end{titlepage}

  \section{Introduction}

  The strong radiative corrections can modify in a significant way the
  effective weak non-leptonic hamiltonian.
  Their effect is studied  using the standard techniques of
   the short-distance expansion
  for the product of two weak currents, improved by the
  renormalization group.
  The analysis of the next-to-leading order
   can give  interesting results, as in  the  decays following the
  $\Delta I = 1/2 $ rule~\cite{Gaillard,Altarelli}
   or in the rare weak B-meson radiative decays~\cite{Grinstein:bsfotone}.

  At more than one loop, the anomalous dimensions
  of the operators of the effective hamiltonian depend on the renormalization
  scheme.
  According to the dimensional regularization method, the UV divergences
  are eliminated by treating the space-time dimension  as a continuous
variable;
  the graphs are  calculated in a  dimension $d$  and  the divergences
  are subtracted
  before taking the limit to the physical dimension $d_p$.
  Beyond the tree level,  a complication  arises from the presence
   of new operators, called
  ``evanescent'',
  i.e. operators that formally disappear when $d \rightarrow d_p$.
  They  can mix with the physical ones and affect the value of the anomalous
  dimension matrix.

  Another source of ambiguity is connected to the way
  a typical four dimension object as
   the $\gamma_5$ matrix is extended to a generic dimension.
  The most used regularization schemes are   Naive  Dimensional
  Regularization (NDR)~\cite{Chanowitz},
   Dimensional Reduction Scheme (DRED)~\cite{Siegel,Altarelli}
  and 't Hooft-Veltman (HV)~\cite{tHooft} scheme:
  in all them  the properties of commutation of $\gamma_5$ with the other
  gamma matrices at $d$ dimension
  are differently fixed.
  We have calculated the next-to-leading QCD correction to the four-fermion
   effective hamiltonian in a new scheme that does not require an
  explicit definition of $\gamma_5$ in $d$ dimension.
  This calculation has been compared with
  analogous calculations~\cite{Buras}  in NDR, DRED and HV;
  our results   explain the
  formal independence
  of the physical results
  from the
  scheme.
  Moreover, in our scheme the problem of evanescent operators can be
  formulated in a particularly simple way.
  It  also has  the  advantage that it can be  easily
  implemented in algebraic computer programs (we used \cite{mathematica}).

  In the first section we describe the general framework to build the
  weak four-fermions effective
  hamiltonian and our approach. In section two,
  three, four and five  we give the details of the calculation of the
  operator anomalous dimension up to two-loops order.
  Section six is devoted to clarify the evanescent
  operators problem. Section seven contains further details
   of our scheme. In section eight we show that our methods  reproduce the
  correct Adler anomaly.

  \section{ Effective Hamiltonian}
  Let us consider a generic four-fermion  weak process
   induced by charged currents.
  In the lowest order in the weak coupling the transition matrix element
  from an initial state I to a final state F is given
  by the time-ordered product
  of the two weak charged currents folded with the W propagator $D_W$:
  \beq
  H_{eff} \simeq g^2_W \int d^4x D_W (x^2, M^2_W)
  <F|T[J^\mu(x)\,J^\dagger_\mu(0)]|I>
  \eeq
  $M_W,\,g_W$ are respectively the $W$ mass and the
  weak coupling constant.

  In the limit $M_W\rightarrow \infty$ let us use the short distance
  operator product expansion for the time-ordered product;
  the $\Delta \, P =1$ effective hamiltonian (P stands for $S, C,$ etc.)
   has the form \cite{Gaillard}:
  \beq
  H_{eff} = \frac{G_F}{\sqrt{2}}\sum_{i} \:C_i\; V^{CKM}_i \,
  \bar{\psi}_4\,X_{4 3}^i\,\psi_3\:  \bar{\psi}_2\,X_{2 1}^i\,\psi_1
  \label{effectivehamiltonian}
  \eeq
  where $C_i$ are the Wilson coefficient functions,
  $V^{CKM}_i$ are appropriate products
  of the Cabibbo-Kobayashi-Maskawa matrix elements,
  $\psi_i$ are the quark fields and $X_{i j}^n$ are the tensor color factors
that include
  color effects.
  The penguin contributions cannot
   arise since we adopt the massless quark limit.
  The mixing with the penguin operators can be avoided in the massive case
  considering four fermions interactions with all different flavours.

  As an example, let us write
  the  charm-changing effective hamiltonian:
  \beqn
  H_{eff} &=& \frac{G_F}{\sqrt{2}} \:
  \left[C_1\,V^{C K M}_1 \,\bar{s}_\alpha \gamma_\mu
  (1-\gamma_5) c_\alpha\:\bar{u}_\beta \gamma_\mu (1-\gamma_5) d_\beta
  + \right.\\
  & & \left. C_2\,V^{C K M}_2\, \bar{u}_\alpha \gamma_\mu
  (1-\gamma_5) c_\alpha\:\bar{s}_\beta \gamma_\mu (1-\gamma_5) d_\beta
  \right]\nonumber
  \eeqn
  $\alpha$ and $\beta$ are color indexes. At the W-mass scale
  $C_1 = 1  $ and $C_2 = 0$.

  Dimensional regularization requires one to deal with the
  generalization of the Dirac $\gamma$-algebra to $d\equiv 4-2\;
  \epsilon$ dimensions; such algebra is known as the Clifford algebra.
  The Clifford algebra in $d$ dimensions is defined by the anticommutation
  relations:
  \beq
  \{\gamma_\mu,\gamma_\nu\}=2 \, \delta_{\mu\nu}\,{\bf{1}} \\
  (1\leq \mu,\nu\leq\,d) \\Tr{\bf{1}} =4
  \label{RelFond}
  \eeq
  where ${\gamma_\mu}$ are the generators, $\bf{1}$ the unit element of the
  algebra and $\delta_{\mu\nu}$ the metric tensor in the euclidean space.
  The space of matrices acting on Dirac spinor indexed is spanned by
    the completely
  antisymmetrical products:
  \beq
  \Gamma^{k} = \Gamma_{\{ \mu_1...\mu_k \}}^k = \frac{1}{k !}\: \sum_{perm} \,
(-1)^p \gamma_{\mu_1}\; ... \;\gamma_{\mu_k}
  \label{CliffordBasis}
  \eeq
  When the number of dimensions is an integer, the total number
  of these matrices is:
  $
  \sum_{j=0}^{d}
  {d \choose j} = 2^d
  $.
   However, when $d$ is noninteger
  the number of matrices must be considered as infinite.

  Let us fix the external fields and omit them in the notation.

  For instance:
  \beq
  \Gamma^{1} \, \otimes \, \Gamma^{1}   \nonumber
  \eeq
   stands for:
  \beq
  \bar{\psi}_4\,\Gamma^{1} \psi_3\:\bar{\psi}_2 \Gamma^{1}\psi_1 \nonumber
  \eeq
  We also define the following operators in the SU(3) color space:
  \beq
  {\bf 1}_{\alpha \,\beta\:,\gamma \,\delta} = \delta_{\alpha
\,\beta}\,\otimes\,
  \delta_{\gamma \,\delta} \\
  {\bf \tilde{1}}_{\alpha \,\beta\:,\gamma \,\delta} = \delta_{\alpha
  \,\delta} \,\otimes\, \delta_{\gamma \,\beta}
  \eeq
  The QCD corrections to the the flavour changing effective
  hamiltonian~(\ref{effectivehamiltonian}) have
  already been evaluated in NDR, DRED and HV  schemes; in all these schemes
  the properties of commutation of $\gamma_5$ at $d$ dimension are stated
  differently, but
  the anomalous dimensions are connected   by a finite
  renormalization and the physical results are the same ~\cite{Buras}.

  We are interested in the renormalization of the operator:
  \beq
  \gamma_\mu \,(1-\gamma_5) \otimes \gamma_\mu \,(1-\gamma_5)
  \label{basictensor}
  \eeq
  It is useful to rewrite operator ~(\ref{basictensor})
  in the form $ E + O$, where $ E$ and $O$
   are respectively
  even and odd under the exchange $\gamma_5 \rightarrow -\gamma_5$:
  \beqn
   E &=&
  1/2 \left[\gamma_\mu \,(1-\gamma_5) \otimes \gamma_\mu \,(1-\gamma_5)
  + \gamma_\mu \,(1+\gamma_5) \otimes \gamma_\mu \,(1+\gamma_5)
  \right] \\
  O &=& 1/2 \left[\gamma_\mu \,(1-\gamma_5) \otimes \gamma_\mu \,(1-\gamma_5)
  - \gamma_\mu \,(1+\gamma_5) \otimes \gamma_\mu \,(1+\gamma_5)
  \right]
  \eeqn
  Since the QCD corrections do not depend on the sign of $\gamma_5$,
  we expect from parity considerations that there is no mixing between
  $ E $ and $ O$ and that the  anomalous dimensions
  of the operators (\ref{basictensor}) and $ E $
  are the same; then
  we can  limit the study to the operator $E$.

  One advantage of this scheme is that there is no need of an explicit
  definition of $ \gamma_5$, so   the equivalence of the
  NDR, DRED and HV schemes emerges trivially.
  Let us introduce the following operators:
  \beqn
  O_1 &=& \Gamma^{1} \, \otimes \, \Gamma^{1}  \nonumber \\
  O_3 &=&\frac{1}{6} \: \Gamma^{3} \, \otimes \, \Gamma^{3} \nonumber \\
  O_5 &=& \Gamma^{5} \, \otimes \, \Gamma^{5} \nonumber \\
  O_M &=& \frac{1}{6} \: \Gamma^{1} \otimes \Gamma^3 \nonumber \\
  {O^\prime}_M &=& \frac{1}{6} \:  \Gamma^3 \otimes \Gamma^{1}
  \label{primabase}
  \eeqn
  They are well defined at a generic dimension $d$. They extend the
  $4$ dimensional operators to
  $d$ dimension; their difference is evanescent,
  that is, when $ d \rightarrow 4$:
  \beqn
  O_1 &\rightarrow& \gamma_\mu \otimes \gamma_\mu \\
  O_3 &\rightarrow& \gamma_\mu\,\gamma_5 \otimes \gamma_\mu\, \gamma_5 \\
  O_5 &\rightarrow& 0 \\
   \frac{1}{6} \left(
  {O_M}_{\{\alpha\, \beta\, \gamma\}\otimes \mu}+
  {{O_M}^\prime}_{\{\alpha\, \beta\, \gamma\}\otimes \mu}
  \right) \epsilon_{\alpha\, \beta\, \gamma\, \mu} &\rightarrow&
   \gamma_\mu
  \otimes \gamma_\mu\,\gamma_5+
   \gamma_\mu \gamma_5
  \otimes \gamma_\mu
  \eeqn
  We will  explicitly show in section (\ref{OMOMPRIME})
  that \{$O_M$, ${O^\prime}_M $\} do not mix with
  \{$ O_1$, $O_3$\}.

  The extension of $E$
  at a generic $d$ dimension   can then be defined as the sum of
   $O_1+ O_3$.
  We will show later that  operator  $O_1+ O_3$
  reproduce the  known physical results.
  We will sometimes use the following  basis:
  \beq
  Q_1 = {\bf 1}\:O_1 \\ Q_2 = \ff \:O_1 \nonumber
  \eeq
  \beq
  Q_3 = {\bf 1}\:O_3 \\ Q_4 = \ff \:O_3 \nonumber
  \eeq
  \beq
  Q_5 = {\bf 1}\:O_5 \\ Q_6 = \ff \:O_5
  \label{basenaive}
  \eeq

   \section{Renormalization Group Analysis}
  The renormalized operators $Q_i$ and the quark-field $\psi$
  can be written in terms of the bare quantities $Q_i^B$ and $\psi^B$ :
  \beq
  Q_i^B = M_{i j}\:Q_j\\
  \psi^B = Z_{\psi}^{1/2} \: \psi
  \eeq
  Let us indicate by ${\it{G}}_i$ and ${\it{G}}_i^B$ the
  finite four-quarks one-particle irreducible green functions
   with the  insertion
  of the operators $Q_i$ and $Q_i^B$ respectively;
  they satisfy the relation:
  \beq
  {\it{G}}_i^B = Z_{i j} \:{\it{G}}_j =  Z_{\psi}^{-2}\,M_{i j}
  \, {\it{G}}_j
  \eeq
  where:
  \beqn
  Z_{\psi} &=& 1+\frac{z_1}{\epsilon}+ \frac{z_2}{\epsilon^2} + ...\\ Z &=&
  1+ a_0+\frac{a_1}{\epsilon}+ \frac{a_2}{\epsilon^2} + ...\\ m &=&
  1+m_0+\frac{m_1}{\epsilon}+ \frac{m_2}{\epsilon^2} + ...\\
  \eeqn
  The lower indexes indicate the powers of
  $1 / \epsilon$.

  Each pole part can
  be expressed as a series
  in the powers of the strong coupling constant $g^2$, for instance:
  \beq
  m_1 =  m_1^{(1)}\: \frac{g^2}{16\,\pi^2} +m_1^{(2)}\:
  \frac{g^4}{(16\,\pi^2)^2} + ...
  \eeq
  The upper indexes indicate the order in $ g^2$.

  In the minimal subtraction scheme, the subtraction of the
   field wave function is performed
  according to the  formulas:
  \beq
  m_1^{(1)} = {a_1}^{(1)} + 2 {z_1}^{(1)} \\
  m_1^{(2)} = {a_1}^{(2)} + 2 {z_1}^{(2)}
  \label{sottrazione}
  \eeq
  where:
  \beqn
  {z_1}^{(1)} &=& -C_f\, \xi
  \nonumber \\
  {z_2}^{(2)} &=& -\frac{17}{4} \,C_a\, C_f +
  \frac{3}{4} \,C_f^2 + \frac{1}{2}\, n_f\,C_f + \,C_a\,C_f
  ( 1-\xi) \left[ \frac{5}{4} -
  \frac{1}{8} (1-\xi) \right]
  \eeqn
  $\xi$ is the gauge parameter and it is $1$ in the Feynman gauge;
  $n_f$ is the number of the quarks running in the fermionic loops.

  Let us call $<Q_i>_N$  the insertions  of the operator $Q_i$
  into all the four-quarks graphs (with the relative counterterms)
   at the N-loops order,
   subtracted according to (\ref{sottrazione}).

  Then
  $m_1^{(N)}$, $m_2^{(N)}$... are the pole parts of $<Q_i>_N$.

  In the regularized $d$-dimensional theory,
  the anomalous dimension matrix $\gamma$  satisfies the
  following renormalization group equation:
  \beq
  ( \partialmu + \beta \frac{\partial}{\partial \, g}\delta_{i j}+
   \gamma_{ij} )\: Q_j = 0
  \eeq
  where:
  \beq
  \gamma = M^{-1} \: \beta(\epsilon , g) \: {\partial \over {\partial g}} M
  \eeq
  Let us fix the notations:
  \beqn
  \epsilon &=& (4-d)/2 \nonumber \\
  \gamma &=& \gamma^{(1)}\: \frac{g^2}{16\,\pi^2} +\gamma^{(2)}\:
  \frac{g^4}{(16\,\pi^2)^2} + ... \nonumber
  \eeqn
  The following relations are well known:
  \beqn
  \beta(\epsilon,g) &=& -\epsilon\:g + \beta(g) \nonumber \\
  \beta(g) &=& -b_0 \,\frac{g^3}{16\,\pi^2} - b_1\,
  \frac{g^5}{(16\,\pi^2)^2}+..\nonumber \\
  b_0 &=& \frac{1}{3}\, ( 11\, C_a -2\, n_f) \nonumber \\
  b_1 &=& \frac{34}{3} \,C_a^2 -\frac{10}{3}\,C_a\,n_f -2 \,C_f\,n_f
  \nonumber
  \eeqn
  In the SU(N) color group:
  \beqn
  C_a &=& N \nonumber \\
  C_f &=& \frac{N^2-1}{2\,N} \nonumber
  \eeqn

  \section{Anomalous Dimension: One Loop}

  Let us define:
  \beq
  O_{+} = \frac{1}{2}\,\left( {\bf 1} + \ff \right) \;
   \left[ \gamma_{\mu} ( 1-\gamma_5) \otimes \gamma_{\mu}( 1-\gamma_5)\right]
  \eeq
  \beq
  O_{-} = \frac{1}{2}\,\left({\bf 1} - \ff \right) \;
   \left[\gamma_{\mu} ( 1-\gamma_5) \otimes \gamma_{\mu}  ( 1-\gamma_5)\right]
  \eeq
  At the one loop
  level the anomalous dimensions are renormalization-scheme independent:
  \beq
  \gamma_{i j}^{(1)} =-
  2\,({m_1}^{(1)})_{i j}
  \eeq

  The anomalous dimensions of $O_{+}$ and $O_{-}$
   are well known~(\cite{Gaillard}):
  \beq
  \gamma_{+} = \frac{\alpha_s}{4 \pi} \,(6-\frac{6}{N}); \\
  \gamma_{-} = \frac{\alpha_s}{4 \pi} \,(-6-\frac{6}{N});
  \eeq

  The mixing of the operators
  $Q_i$ gives the following results:
  \beqn
  <Q_1>_1 &\rightarrow&   \left(\frac{3}{C_a} \,Q_3 - 3
  \, Q_4 \right)\,\frac{1}{\epsilon} \nonumber \\
  <Q_2>_1 &\rightarrow& \left[-\frac{3}{2}
  \,( Q_1+Q_3) +\frac{3}{2} \,C_a \,Q_2 + (
  -\frac{3}{2}\,C_a +\frac{3}{C_a}) \,Q_4 \right]
  \,\frac{1}{\epsilon} \nonumber \\
  <Q_3>_1 &\rightarrow&  \left[\frac{3}{C_a} \,Q_1 - 3\, Q_2 +
   \frac{1}{12\,C_a} \,Q_5 -\frac{1}{12}\, Q_6 \right] \,\frac{1}{\epsilon}
  \nonumber \\
  <Q_4>_1 &\rightarrow&
  \left[-\frac{3}{2} \,
  ( Q_1+Q_3)  + \left(-\frac{3}{2}\,C_a +\frac{3}{C_a}\right) \,Q_2+
    \right. \nonumber\\
  & & \left.
  +\frac{3}{2} \,C_a \,Q_4
   -\frac{1}{24}\,Q_5+ \left( -\frac{C_a}{24} +\frac{1}{12\,C_a}
   \right)\,\,Q_6\right]\,\frac{1}{\epsilon} \nonumber \\
  <Q_5>_1 &\rightarrow& -\frac{120}{C_a}\: Q_3  +
   120\:Q_4 +16\, C_f\,\frac{1}{\epsilon}\:Q_5 + ...\nonumber \\
  <Q_6>_1 &\rightarrow& 60\: Q_3 + \left(
    60\, C_a - \frac{120}{C_a} \right)\: Q_4 + \nonumber\\
  & & \left[
  \frac{21}{2} \,Q_5+ \left( 16\, C_f -\frac{21}{2}\,C_a \right)\, Q_6
  \right]\:\frac{1}{\epsilon} + ...
  \label{operatori::1loop}
  \eeqn

  Only the operators $Q_1,\,Q_2,\,Q_3,\,Q_4$ have
  non-zero projections at four dimensions.

  The anomalous dimension matrix is:
  \beq
  \gamma^{(1)} =\left(
  \matrix{
   0 & 0 & -\frac{6}{N} & 6  & 0 & 0 \cr
   3 & -3\,N & 3 & 3\,N-\frac{6}{N} & 0 & 0  \cr
   -\frac{6}{N} & 6 & 0 & 0 & -\frac{1}{6\,N} & \frac{1}{6} \cr
   3 & 3\,N-\frac{6}{N} & 3 & -3\, N & \frac{1}{12} & \frac{N}{12}
  -\frac{1}{6 N} \cr
  0 & 0 & 0 & 0 & -32 \, C_f & 0 \cr
  0 & 0 & 0 & 0 & -21 & -32\,C_f+21\,N
   } \right)
  \label{mixing::loop1}
  \eeq
  Now let us change basis and choose operators that diagonalize the physical
  sector of the
   anomalous dimension matrix (\ref{mixing::loop1}),
  in order to simplify the calculation and make  easier the comparison
  with the previous literature (f.i. ~\cite{Buras}):
  \beqn
  Q_1^{\prime} &=& \frac{1}{2} \: [\,(Q_1 + Q_3)+(Q_2+Q_4)\,] \nonumber \\
  Q_2^{\prime} &=& \frac{1}{2} \: [\,(Q_1 + Q_3)-(Q_2+Q_4)\,] \nonumber \\
  Q_3^{\prime} &=&  -Q_2 + Q_4 \nonumber \\
  Q_4^{\prime} &=& \frac{1}{N} \: [\,N\,(Q_3 - Q_1)+(Q_2-Q_4)\,] \nonumber \\
  Q_5^{\prime} &=& Q_5 \nonumber \\
  Q_6^{\prime} &=& Q_6
  \label{Oprime}
  \eeqn

  In the basis $Q_i^\prime$ we have:
  \beq
  {{\gamma}^{\prime}}^{(1)} = \frac{\alpha_s}{4 \pi}\: \left(
  \matrix{
   6-\frac{6}{N} & 0 & 0 & 0  & \frac{1}{24}-\frac{1}{12\,N} &
  \frac{1}{12}\,\left(1-\frac{1}{N}+3\,N\right) \cr
   0 & -6-\frac{6}{N} & 0 & 0 & -\frac{1}{24}-\frac{1}{12\,N} &
  \frac{1}{12}\,\left(1+\frac{1}{N}-3\,N\right) \cr
   0 & 0 &  -6\,N+\frac{6}{N} & 0  & \frac{1}{12} &
  \frac{N}{2} -\frac{1}{6\,N}\cr
   0 & 0 & 0 & \frac{6}{N} & -\frac{1}{4\,N } & -\frac{1}{3} +
  \frac{1}{6\,N^2}  \cr
  0 & 0 & 0 & 0 & -32 \, C_f & 0 \cr
  0 & 0 & 0 & 0 & -21 & -32\,C_f+21\,N }
  \right)
  \eeq

  Let us notice the agreement of the
   anomalous dimensions of $Q_1^\prime$ and $Q_2^\prime$ with
   $O_{+}$ and $O_{-}$.

  At one loop, $O_{+}$, $O_{-}$  and $Q_1^\prime$, $Q_2^\prime$
  have the same pole parts but the finite terms are different;
  they are  equalized by the following finite renormalization:
  \beq
  O_{+} = Q_1^\prime + \frac{g^2}{16\,\pi^2}\, p_j^+ \, Q_j^\prime \nonumber \\
  O_{-} = Q_2^\prime + \frac{g^2}{16\,\pi^2}\, p_j^- \, Q_j^\prime
  \label{inserzioni::finite}
  \eeq
  where:
  \beqn
  p^+_{1} &=& \frac{11}{2\,N} -\frac{33}{8} -\frac{11}{8}\,N \nonumber\\
  p^+_2 &=& -\frac{5}{8} \; (N+1) \nonumber\\
  p^+_{3} &=& \frac{1}{4}-\frac{1}{4\,N^2}
  -\frac{9}{8\,N} +\frac{9\,N}{8}  \nonumber \\
  p^+_{4} &=& -\frac{7}{8} -N -\frac{1}{4\,N}\nonumber \\
  p^-_{1} &=& -\frac{5}{8} \; (N-1)\nonumber \\
  p^-_2 &=&  \frac{11}{2\,N} +\frac{33}{8} -\frac{11}{8}\,N \nonumber \\
  p^-_{3} &=&  \frac{1}{4}-\frac{1}{4\,N^2} +\frac{9}{8\,N} -\frac{9\,N}{8}
  \nonumber \\
  p^-_{4} &=& \frac{7}{8} -N -\frac{1}{4\,N}
  \label{elencop}
  \eeqn
  Let us notice that
   we can  neglect the terms proportional to $Q_5^\prime$ and
  $Q_6^\prime$ without any physical consequence, as they are evanescent
operators.

  \section{$O_M$, ${O^\prime}_M$ Insertions}
  \label{OMOMPRIME}

  Let us show that operators of the form $\Gamma \otimes \tilde{\Gamma}$
  do not mix with the operators $\Gamma \otimes \Gamma$,
  $\tilde{\Gamma} \otimes \tilde{\Gamma}$; then their anomalous dimensions
  do not affect the
  evolution of the coefficients of the effective  hamiltonian.

  $\Gamma$ and $\tilde{\Gamma}$  stand indifferently for   $\Gamma^1$ or
  $\Gamma^3$.

  A generic one-loop graph with  $\Gamma \otimes \tilde{\Gamma}$ insertion
  can be considered as the product of:
  \begin{description}
  \item[(A)] a color factor
  \item[(B)] a Dirac tensor structure coming from the fermionic lines (4 $
  \gamma$ matrices plus $\Gamma$ and $\tilde{\Gamma}$)
  \item[(C)] a Dirac tensor structure coming
  from the loop integration.
  \end{description}

  The product of the local part of (C) with (B)  can give rise only to the
  following five tensor structures, where  $\alpha$, $\beta$,
  $\alpha^\prime$ and $\beta^\prime$ are contracted in all possible ways:
  \beqn
  && \gamma_\alpha\;\gamma_\beta\;\Gamma\;\gamma_{\alpha^\prime }
  \;\gamma_{\beta^\prime} \otimes   \tilde{\Gamma}
  \label{(1)}\\
  && \Gamma \gamma_\alpha\gamma_\beta \otimes \tilde{\Gamma}
  \gamma_{\alpha^\prime}
  \gamma_{\beta^\prime}
  \label{(2)}\\
  && \gamma_\alpha\gamma_\beta\Gamma
  \otimes\gamma_{\alpha^\prime}\gamma_{\beta^\prime}
  \tilde{\Gamma}
  \label{(3)} \\
  &&\Gamma \gamma_\alpha\gamma_\beta \otimes
   \gamma_{\alpha^\prime}\gamma_{\beta^\prime}
  \tilde{\Gamma}
  \label{(4)}\\
  &&\gamma_\alpha\gamma_\beta\Gamma \otimes \tilde{\Gamma}
  {\gamma_\alpha}^\prime
  {\gamma_\beta}^\prime
  \label{(5)}\\
  \eeqn
  Let us fix the contraction of (\ref{(1)}) in this way:
  \beq
  \gamma_\alpha\;\gamma_\beta\;\Gamma\;\gamma_\beta
  \;\gamma_\alpha \otimes   \tilde{\Gamma}
  \label{1contrazione}
  \eeq
  It is obvious that this tensor product is proportional to
  $\Gamma \otimes \tilde{\Gamma}$ as
  \beq
  \gamma_\mu\;\Gamma^{N}\;\gamma_\mu = (-1)^N \,(d-2\,N)\;\Gamma^{N}
  \eeq
  Any other possible contractions can be reduced to (\ref{1contrazione})
  without modifying the $\Gamma \otimes \tilde{\Gamma}$ structure, according
  to  (\ref{RelFond}).

  Now let us turn to (\ref{(2)}) and (\ref{(3)}) and fix $\Gamma =
  \Gamma^1$,
  $\tilde{\Gamma}=\Gamma^3$, $\alpha = \alpha^\prime$,
  $\beta = \beta^\prime$:
  \beqn
  &&{\Gamma^1}_\mu \gamma_\alpha\gamma_\beta \otimes
{\Gamma^3}_{\{\nu\,\rho\,\sigma\}}
  \gamma_{\alpha} \gamma_{\beta}  \: \epsilon_{\nu\rho\sigma\mu}
   \rightarrow \nonumber \\
  &&
  \left[ 6\, {\Gamma^3}_{\{\alpha\,\beta\,\gamma\}} \otimes
  {\Gamma^1}_\xi
  + (6+d) \,{\Gamma^1}_\xi \otimes  {\Gamma^3}_{\{\alpha\,\beta\,\gamma\}} +
6\,
   \Gamma^3 \otimes
  \Gamma^3 \right] \epsilon_{\alpha\beta\gamma\xi} \label{pizza1} \\
  &&
  \gamma_\alpha\gamma_\beta \Gamma^1_\mu
  \otimes\gamma_{\alpha}\gamma_{\beta}
  \Gamma^3_{\{\nu\,\rho\,\sigma\}} \: \epsilon_{\nu\rho\sigma\mu}
    \rightarrow \nonumber \\
  &&\left[6 \,{\Gamma^3}_{\{\alpha\,\beta\,\gamma\}} \otimes
  {\Gamma^1}_\xi
  + (6+d)\, {\Gamma^1}_\xi \otimes  {\Gamma^3}_{\{\alpha\,\beta\,\gamma\}} -
  6\, \Gamma^3 \otimes \Gamma^3 \right]
  \epsilon_{\alpha\beta\gamma\xi}
  \label{pizza2}
  \eeqn
  Using (\ref{RelFond}) and the Fierz properties in the Dirac space,
   the product tensor has been  reduced to a simpler form;
  the free indexes have been contracted
   with the completely 4-dimension antisymmetric tensor
  $\epsilon_{\alpha\beta\gamma\xi}$.

  We are interested in pole parts; they are independent from the
  external momenta (but be careful they are able to
    regularizes both the UV than the IR
  divergences in the massless limit), so for each graph
  we are free to change the inner and outer momenta
  in order to have the same (A) and (C) to multiply
  (\ref{(2)}) and (\ref{(3)}).
  In the sum only the terms $\Gamma \otimes \tilde{\Gamma} $ survive.
  Different contractions of $\alpha\,\beta\,\alpha^\prime\,\beta^\prime$
  are related to (\ref{pizza1}), (\ref{pizza2})  using (\ref{RelFond}); they
  cannot restore the terms
  $\Gamma \otimes \Gamma $.

  At the same way:
  \beqn
  &&{\Gamma^1}_\mu \gamma_\alpha\gamma_\beta \otimes
   \gamma_\beta\gamma_\alpha  {\Gamma^3}_{\{\nu\,\rho\,\sigma\}} \;
   \epsilon_{\nu\rho\sigma\mu}
   \rightarrow  \nonumber \\
  &&
   \left[-6\, {\Gamma^3}_{\{\nu\,\rho\,\sigma\}} \otimes
  {\Gamma^1}_\mu
  + (6+d) {\Gamma^1}_\mu \otimes  {\Gamma^3}_{\{\nu\,\rho\,\sigma\}} - 6\,
\Gamma^3 \otimes
  \Gamma^3 \right] \epsilon_{\nu\rho\sigma\mu}
  \\
  &&\gamma_\alpha\gamma_\beta{\Gamma^1}_\mu \otimes
{\Gamma^3}_{\{\nu\,\rho\,\sigma\}}
  \gamma_\beta
  \gamma_\alpha  \;
   \epsilon_{\nu\rho\sigma\mu} \rightarrow  \nonumber \\
  && \left[-6\, {\Gamma^3}_{\{\nu\,\rho\,\sigma\}} \otimes
  {\Gamma^1}_\mu
  + (6+d)\, {\Gamma^1}_\mu \otimes  {\Gamma^3}_{\{\nu\,\rho\,\sigma\}} + 6\,
\Gamma^3 \otimes
  \Gamma^3\right] \epsilon_{\nu\rho\sigma\mu}
  \eeqn
  At two loops order we can repeat the same reasonings.

  Now it is obvious that graphs (23-16-6-12-21-25) in figure 3  are of the
  form $\Gamma \otimes \tilde{\Gamma}$, as well as
  graphs (26-27).

  The other graphs can be divided into two classes:

  (a) one fermionic line has only one gluon attached;

  (b) both fermionic line have at least 2 gluons attached;

  Class (a) operators are reducible to the
  tensor products already examined plus:
  \beqn
  &&
  \gamma_\alpha\Gamma\gamma_\beta
  \otimes \gamma_{\alpha}\tilde{\Gamma}\gamma_{\beta} \\
  &&
  \gamma_\alpha\Gamma\gamma_\beta
  \otimes \gamma_{\alpha}\gamma_{\beta}\tilde{\Gamma}\\
  &&
  \gamma_\alpha\Gamma\gamma_\beta
  \otimes \tilde{\Gamma}\gamma_{\alpha}\gamma_{\beta}
  \eeqn
  They can be simplified according to (\ref{RelFond}); each one gives only
terms
  proportional
  to $\Gamma \otimes \tilde{\Gamma}$.
  Class (b) operators give rise to  the additional products:
  \beqn
  &&
  \Gamma\gamma_\alpha\gamma_\beta\gamma_\rho\gamma_\sigma \otimes
  \tilde{\Gamma}\gamma_{\alpha} \gamma_{\beta}
  \gamma_{\rho}\gamma_{\sigma}
  \label{2loop1}\\
  &&
  \gamma_\alpha\gamma_\beta\gamma_\rho\gamma_\sigma \Gamma\otimes
  \gamma_{\alpha} \gamma_{\beta}
  \gamma_{\rho}\gamma_{\sigma} \tilde{\Gamma}
  \label{2loop2}\\
  &&
  \gamma_\alpha\gamma_\beta\gamma_\rho\gamma_\sigma \Gamma\otimes
  \tilde{\Gamma}
  \gamma_{\alpha} \gamma_{\beta}
  \gamma_{\rho}\gamma_{\sigma}
  \label{2loop3}\\
  &&
  \Gamma\gamma_\alpha\gamma_\beta\gamma_\rho\gamma_\sigma \otimes
  \gamma_{\alpha} \gamma_{\beta}
  \gamma_{\rho}\gamma_{\sigma} \tilde{\Gamma}
  \label{2loop4}\\
  &&
  \gamma_\alpha\gamma_\beta\Gamma\gamma_\rho\gamma_\sigma \otimes
  \tilde{\Gamma} \gamma_{\alpha} \gamma_{\beta}
  \gamma_{\rho}\gamma_{\sigma}
  \label{2loop5}\\
  &&
  \gamma_\alpha\gamma_\beta\Gamma\gamma_\rho\gamma_\sigma \otimes
  \gamma_{\alpha} \gamma_{\beta} \tilde{\Gamma}
  \gamma_{\rho}\gamma_{\sigma}
  \label{2loop6}\\
  &&
  \gamma_\alpha\gamma_\beta \Gamma \gamma_\rho\gamma_\sigma \otimes
  \gamma_{\alpha} \gamma_{\beta}
  \gamma_{\rho}\gamma_{\sigma} \tilde{\Gamma}
  \label{2loop7}
  \eeqn
  The tensor products (\ref{2loop1})-(\ref{2loop2})
   and (\ref{2loop3})-(\ref{2loop4}) give terms $\Gamma \otimes \Gamma$
  that cancel each other; the structures
   (\ref{2loop5})-(\ref{2loop6})-(\ref{2loop7})
   give rise only to terms of the
  form
  $\Gamma \otimes \tilde{\Gamma}$ and $\tilde{\Gamma} \otimes \Gamma $.

  \section{Evanescent Operators}

  At $\alpha^2$ order, the anomalous dimension matrix
  depends on the renormalization scheme; moreover, its value is affected
  by the presence of the evanescent operators that mix with
  the physical ones.

  We can also adopt a non-minimal subtraction scheme where
  the renormalized evanescent operators mix exclusively among
  them; in this scheme,
  they can be suppressed from the beginning in the effective hamiltonian.

  In the minimal subtraction scheme,
   the two-loops anomalous dimensions
  are given by:
  \beq
  (\gamma_{i j}^{(2)})_{MS} = -4 \,{(m_1^{(2)})}_{i j}
  \eeq
  The evanescent operators $Q_5$ and $Q_6$ give extra finite contributions
  of order $\alpha^2$ to the mixing of $Q_3$ (cfr. eq.
  (\ref{operatori::1loop})).
  Their effect is expressed by the following relation:
  \beq
  \gamma_{i j}^{(2)} \rightarrow
   (\gamma_{i j}^{(2)})_{MS} + \sum_{ev=5,6} \gamma_{i\, ev}^{(1)}\,
  r_{ev\, j}  \\
  {i,j} \in \{1,2,3,4\}
  \label{dimensioneanomala::MS}
  \eeq
  $r_{i j}$ is the reduction matrix of the evanescent operators over the
  physical ones (see for instance \cite{Bonneau1,Bonneau2} ).

  In the non-minimal subtraction scheme,
  the anomalous dimension matrix
  is given by:
  \beq
  \gamma_{i j}^{(2)} = (\gamma_{i j}^{(2)})_{M S}-2 \,b_0 \,{m_0^{(1)}}_{i j}+
  2\,{(m_0^{(1)}\:m_{1\, M S}^{(1)}- m_{1\, M S}^{(1)}\:m_{0}^{(1)})}_{i j}
  \label{dimensioneanomala::NMS}
  \eeq
  Let us choose as finite parts
  just the one-loop projections of the evanescent operators
  :
  \beq
  (m_0^{(1)})_{5 3} = -120\,\frac{1}{C_a} \\ (m_0^{(1)})_{5 4} = 120
  \nonumber
  \eeq
  \beq
  (m_0^{(1)})_{6 3} = 60 \\ (m_0^{(1)})_{6 4} = 60 \,( 4\, C_f -C_a)
  \label{dimensioneanomala::costanti}
  \eeq
  The other matrix elements of $m_0$ are zero.

  In this non-minimal
   prescription we have no mixing among physical and evanescent operators
  in the RG equation for the weak hamiltonian
  coefficients (the anomalous dimension for the coefficients is the transpose
   of $\gamma$); in fact, we obtain:
  \beq
  \gamma_{i j}^{(2)} =0\:\: i\in \{5,6\}\:\:j\in \{1,2,3,4\}
  \label{evanescenti}
  \eeq

  The (\ref{dimensioneanomala::NMS})
   is equivalent to (\ref{dimensioneanomala::MS}) by
  identifying:
  \beq
  \gamma^{(1)} = -2\, m_1^{(1)} \\ r^{(1)} =  m_0^{(1)}
  \eeq

  \section{Two-loops Anomalous Dimension}

  After having eliminated the evanescent operators as described in the
  previous section, we  change basis from $Q_i$ to $Q_i^\prime$.
  Relation (\ref{evanescenti}) is still valid  in the basis $Q_i^\prime$.

  We are working in the non-minimal subtraction scheme
  defined in the preceding section and
  we  compute the anomalous  dimension matrix (shown in Appendix 1) according
to
  eqns. (\ref{dimensioneanomala::NMS}-\ref{dimensioneanomala::costanti}).

  Now let us consider a generic
   finite renormalization from one set of operators $X$ to another set
$X^\star$:
  \beq
  {X^\star}_i = (\delta_{i j} + \frac{g^2}{16\,\pi^2}\, p_{i j})\: X_j
  \eeq
  At order $g^4$ of the perturbative expansion, their anomalous dimensions
  are connected by
  the following relation:
  \beq
  {\gamma_{i j}}^{(2)} =
  {\gamma^\star_{i j}}^{(2)} \,\delta_{i j} -
   (2 b_0 -\gamma_i^{\star\,(1)} +
  \gamma_j^{(1)})\; p_{i\,j}
  \eeq
  Repeated indexes are not summed; we are assuming that
  $\gamma^{(1)},~\gamma^{\star (1)}$ are diagonal matrices.

  The relation (\ref{inserzioni::finite}) implies:
  \beqn
  {\gamma^\prime_{1 j}}^{(2)} &=&
   (-2 b_0 - \gamma_j^{\prime\,(1)} +
  \gamma_+^{(1)})\; p^{+}_j  \hspace{1 in} j \neq 1 \\
  {\gamma^\prime_{2 j}}^{(2)} &=&
   (-2 b_0 - \gamma_j^{\prime\,(1)} +
  \gamma_-^{(1)})\; p^{-}_j  \hspace{1 in} j \neq 2 \\
  p^{+}_{1} &=& \frac{\gamma_{+}^{(2)}-\gamma_{1 1}^{(2)}}{2\,b_0}\\
  p^{-}_{2} &=& \frac{\gamma_{-}^{(2)}-\gamma_{2 2}^{(2)}}{2\,b_0}
  \eeqn
  $\gamma_{+}^{(2)}$ and $\gamma_{-}^{(2)}$ are given in \cite{Buras}.
  Inserting ${\gamma^\prime_{i j}}^{(2)} $ in all these relations,
   we obtain the same values for  $p^+$ and $p^-$ that we had
  from the one-loop finite renormalization
  (\ref{elencop}), as we expected.

  \section{The Adler Anomaly}

  Anomalies can arise when classical symmetries are not preserved after
  regularization.
  Chiral symmetry is valid in the physical space-time
  dimension,
  but not in an arbitrary number of dimensions. This symmetry breaking
  leads to the Adler
  anomaly, responsible of a non-zero  $\pi^0
  \rightarrow 2\; \gamma$ decay rate in the current algebra context.

  The simplest one-loop  correction to the QED decay of the axial current
  $\bar{\psi}\;\gamma_\mu\;\gamma_5\;\psi$ (or of its divergence)
   into two gauge fields
  is given by the sum of the so called ``triangle'' diagrams (fig.
  1).
  The computation of the Adler anomaly
  in dimensional regularization presents again  problems connected with
  the definition of
  $\gamma_5$; it is well known that
  a $\gamma_5$ simply anticommutation with all $\gamma$
  leads to inconsistencies when performing the trace of $\gamma_5$
  with a product of the $\gamma$.

  According to HV scheme, the presence of $\gamma_5$ gives rise to a
  coefficient proportional to $4-d$ that cancels
  the poles of the divergent integrals;
   a non-zero   anomaly results at four dimensions.
  A simpler method is to perform the trace with
  the  vertex
  $1/6\;\Gamma^3$, well-defined at an arbitrary dimension;
  then we  integrate and, if necessary, renormalize. At the end we
  return to four dimensions and
  multiply the results for the completely antisymmetrical tensor at four
  indexes; that is, we
  perform the four dimension limit
   $1/6\;\Gamma^3_{\alpha \beta \gamma} \; \epsilon_{\alpha \beta \gamma
  \sigma} \rightarrow \gamma_5\;\gamma_\sigma $ (a sum  over repeated indexes
  is implied).

  Now let us  calculate the divergence of the total chiral current
   coupled to two photon lines, in the limit of massless quarks.
  A straightforward application of the
  Feynman rules to the  diagrams in fig. (1) yields the expression:
  \beqn
  \hat{q} &\equiv&  q_\omega\,\gamma_\omega \\
  T_{a b c \mu\nu} &=& e^2 \int \frac{d^{d}l}{(2\,\pi)^{d}}\;
  \frac{1}{l^2\,(l-q)^2\,(l+p)^2}\; \\
  & &
  Tr[\Gamma^3_{\{a b c\}}\,(\hat{l}+\hat{p})\,\gamma_\mu\,\hat{l}\,\gamma_\nu\,
  (\hat{l}-\hat{q}) -
  \Gamma^3_{\{a b c\}}\,(\hat{l}-\hat{q})\,\gamma_\nu\,\hat{l}\,\gamma_\mu\,
  (\hat{l}+\hat{p})]
  \eeqn
  The trace multiplied by $\epsilon_{a b c \sigma}\;(p_\sigma + q_\sigma)$
  results to be:
  \beqn
  & 8 & \;\{\epsilon_{\alpha \beta  \mu\nu}\;[ l_\alpha\,(l^2+p\cdot q)
  \,(p_\beta+q_\beta)+ (2\, l^2+ l\cdot p-l\cdot q) \,p_\alpha\,q_\beta]+ \\
   &+& 2\,(l_\nu \epsilon_{\alpha \beta  \rho\mu} -
  l_\mu \epsilon_{\alpha \beta  \rho\nu})\; l_\alpha \,p_\beta\,q_\rho + \\
  &+& ( q_\mu -p_\mu)\, l_\alpha \,p_\beta\,q_\rho
  \epsilon_{\alpha \beta  \rho\nu}+
   ( q_\nu -p_\nu)\, l_\alpha \,p_\beta\,q_\rho
  \epsilon_{\alpha \beta  \rho\mu}\}
  \eeqn
  The four types of loop integrals that appears are easily evaluated
  on-shell ($q^2 = p^2 =0$). In the sum divergent and non-local parts
   are canceled.
  At the end, we find the expected anomaly for the divergence
  of  the axial current:
  \beq
  \epsilon_{a b c \sigma} \;  T_{a b c \mu\nu}
  \; (p_\sigma+q_\sigma) = - \frac{e^2}{2\pi^2} \epsilon_{\alpha \beta
  \mu\nu} \,
   \,p_\alpha\,q_\beta
  \eeq
  At the same way we can find  the one-loop correction to the
  $\pi^0 \rightarrow\;2\,\gamma$ decay in the $\sigma$ model, by substituting
  in the graphs (1)  the vertex $1/24
  \;\gamma_\sigma \epsilon_{\alpha \beta \gamma \sigma} \; \Gamma_{\{\alpha
  \beta \gamma\}}$ to the vertex $\gamma_5$.
  Also in this case
  the amplitude has the expected value, that is:
  \beq
  {\it{A}}(k^2=0) = \frac{e^2\,g}{4\pi^2\,m} \epsilon_{\mu\nu\rho\sigma}
  \epsilon_1^{\mu}\,
  \epsilon_2^{\nu}\,
  p^\rho\;q^\sigma
  \eeq
  where $\epsilon_1^{\mu},
  \epsilon_2^{\nu}$ are
  the polarizations of the photons of momenta $p$ and $q$ $(p^2 =q^2 =0)$,
  $k =q+p $ and $m$ is the mass
  of the inner quark.

  \section{Acknowledgements}

  G.~R.~would like to commend the Physics Department of Harvard for the
  generous hospitality and friendly working atmosphere. She also wants to
  thank Professor G.~Paffuti and Dr.~A.~Vicer\'e of the Physics Department of
  the University of Pisa for useful discussions.  G.~R.~and G.~C.~would both
  like to thank Professor Howard Georgi of Harvard for his careful  reading
  of the manuscript and his constructive suggestions.

  \newpage
  \bigskip
  {\bf Figures and Table Captions}
  \par\noindent
  \begin{description}
  \item [Fig.\ 1] One-loop diagrams  for the Adler anomaly.
  \item [Fig.\ 2] One loop diagrams for the anomalous dimension
  of the four fermion operators.
  \item [Fig.\ 3]
  In fig (3) the $28$ independent two loop diagrams for the
  anomalous dimension of the four fermion operators are listed.
  In the graphs $25,26,27$ the blobs represent the
  gluon self energy; so we need to consider the color factors in the sum and
  $25^1,26^1,27^1$ refer to the insertion of $ 1 \otimes 1 $, while
  $25^2,26^2,27^2$ refer to the insertion of $T^a \otimes T^a$.
  \item [Tab. \ 1]
  The first item numbers the graphs listed in fig. (3);
  the second is the simple pole resulting from the insertions  of the generic
  operator $\Gamma^n \otimes\Gamma^n$ in the graph.
  The gauge parameter $\chi$ stands for  $ 1-\xi$, where
  \[
  {\cal L}_{\rm gauge-fixing} = \frac{1}{2\xi} \left(\partial_\mu
  G_\mu\right)^2
  \]
  is the gauge fixing term, so  $\chi = 0$ in the Feynman gauge; $\rm n_f$ is
the
  number of quark flavours and N refers to the SU(N) color group.
  Multiplicities and color factors of the graphs are not reported
  (are the same than \cite{Altarelli,Buras}).
  \end{description}
  \newpage
  \newpage
  \newpage
  \include{tabelle}
  \newpage
  \vskip3truecm
  \centerline{\bf{Appendix 1: Two-Loop Anomalous Dimension Matrix}}
  \vskip2truecm

  The elements of the two-loop anomalous dimension (see Sect. 7) are:
  \beqn
  {{{\gamma}^{\prime}}^{(2)}}_{1\,1} &=&
    -{{143}\over 3} - {{57}\over {2\,{{{\rm N}}^2}}} + {{39}\over {{\rm N}}} +
     {{325\,{\rm N}}\over {12}} + {{121\,{{{\rm N}}^2}}\over {12}} \nonumber
\\
  {{{\gamma}^{\prime}}^{(2)}}_{1\,2} &=&
    -{{15}\over 2} - {{35\,{\rm N}}\over {12}} + {{55\,{{{\rm N}}^2}}\over
  {12}} \nonumber \\
  {{{\gamma}^{\prime}}^{(2)}}_{1\,3} &=&
    -{{21}\over 2} + {3\over {{{{\rm N}}^3}}} + {{12}\over {{{{\rm N}}^2}}} -
     {{113}\over {12\,{\rm N}}} + {{77\,{\rm N}}\over {12}} -
     {{3\,{{{\rm N}}^2}}\over 2} \nonumber \\
  {{{\gamma}^{\prime}}^{(2)}}_{1\,4} &=&
    {{103}\over {12}} + {3\over {{{{\rm N}}^2}}} + {9\over {{\rm N}}} +
     {{5\,{\rm N}}\over {12}} + {{22\,{{{\rm N}}^2}}\over 3}
  	 \nonumber \\
  {{{\gamma}^{\prime}}^{(2)}}_{2\,1} &=&
    -{{15}\over 2} + {{35\,{\rm N}}\over {12}} + {{55\,{{{\rm N}}^2}}\over
  {12}} \nonumber \\
  {{{\gamma}^{\prime}}^{(2)}}_{2\,2} &=&
    -{{143}\over 3} - {{57}\over {2\,{{{\rm N}}^2}}} - {{39}\over {{\rm N}}} -
     {{325\,{\rm N}}\over {12}} + {{121\,{{{\rm N}}^2}}\over {12}} \nonumber \\
  {{{\gamma}^{\prime}}^{(2)}}_{2\,3} &=&
  {{21}\over 2} + {3\over {{{{\rm N}}^3}}} - {{12}\over {{{{\rm N}}^2}}} -
  	{{113}\over {12\,{\rm N}}} + {{77\,{\rm N}}\over {12}} +
     {{3\,{{{\rm N}}^2}}\over 2} \nonumber \\
  {{{\gamma}^{\prime}}^{(2)}}_{2\,4} &=&
    {{103}\over {12}} + {3\over {{{{\rm N}}^2}}} - {9\over {{\rm N}}} -
     {{5\,{\rm N}}\over {12}} + {{22\,{{{\rm N}}^2}}\over 3}
   \nonumber \\
  {{{\gamma}^{\prime}}^{(2)}}_{3\,1} &=&
    -{{75}\over 2} + {{54}\over {{{{\rm N}}^2}}} - {{78}\over {{\rm N}}} +
     {{349\,{\rm N}}\over 6} + {{10\,{{{\rm N}}^2}}\over 3} \nonumber \\
  {{{\gamma}^{\prime}}^{(2)}}_{3\,2} &=&
    {{75}\over 2} - {{54}\over {{{{\rm N}}^2}}} - {{78}\over {{\rm N}}} +
     {{349\,{\rm N}}\over 6} - {{10\,{{{\rm N}}^2}}\over 3}
  \nonumber \\
  {{{\gamma}^{\prime}}^{(2)}}_{3\,3} &=&
    {{85}\over 2} - {{21}\over {2\,{{{\rm N}}^2}}} - 32\,{{{\rm N}}^2}
  \nonumber \\
  {{{\gamma}^{\prime}}^{(2)}}_{3\,4} &=&
   {{-18}\over {{\rm N}}} + {{127\,{\rm N}}\over 6}
  \nonumber \\
  {{{\gamma}^{\prime}}^{(2)}}_{4\,1} &=&
    -{{224}\over 3} - {{54}\over {{{{\rm N}}^3}}} +
     {{132}\over {{{{\rm N}}^2}}} - {{93}\over {2\,{\rm N}}} +
     {{57\,{\rm N}}\over 2} + {{44\,{{{\rm N}}^2}}\over 3}
  \nonumber \\
  {{{\gamma}^{\prime}}^{(2)}}_{4\,2} &=&
    -{{224}\over 3} + {{54}\over {{{{\rm N}}^3}}} +
     {{132}\over {{{{\rm N}}^2}}} + {{93}\over {2\,{\rm N}}} -
     {{57\,{\rm N}}\over 2} + {{44\,{{{\rm N}}^2}}\over 3}
  \nonumber \\
  {{{\gamma}^{\prime}}^{(2)}}_{4\,3} &=&
    {{18}\over {{{{\rm N}}^3}}} - {{145}\over {6\,{\rm N}}} +
     {{37\,{\rm N}}\over 6} \nonumber \\
  {{{\gamma}^{\prime}}^{(2)}}_{4\,4} &=&
    -{{35}\over 6} + {{51}\over {2\,{{{\rm N}}^2}}} +
     {{44\,{{{\rm N}}^2}}\over 3}
  \nonumber
  \eeqn
  A common factor
  $\frac{{\alpha_s}^2}{{4 \pi}^2} $ is subtended.